\documentclass[twocolumn]{jpsj3}
\usepackage{times}
\usepackage[dvips]{color}

\def\lsim{\lower -0.3ex \hbox{$<$} \kern -0.75em \lower 0.7ex \hbox{$\sim$}}
\def\gsim{\mathrel{\rlap{\lower4pt\hbox{\hskip1pt$\sim$}}
    \raise1pt\hbox{$>$}}}                

\def\tn{$T_{\rm AF}$}
\def\jo #1#2#3#4{#1 {\bf #2} (#3) #4}

\def\PRB{Phys.\ Rev.\ B}

\def\JCP{J.\ Chem.\ Phys.}

\def\JPSJ{J.\ Phys.\ Soc.\ Jpn.}

\def\JPP{J.\ Phys.\ (Paris)}

\def\CR{Chem.\ Rev.}

\def\JACS{J.\ Am.\ Chem.\ Soc.}
\def\CL{Chem.\ Lett.}
\def\IC{Inorg.\ Chem.}
\def\JMC{J.\ Mater.\ Chem.}
\def\CPL{Chem.\ Phys.\ Lett.}

\usepackage{bm} 

\usepackage[normalem]{ulem}

\usepackage{color}

\usepackage{ulem}

\title{
Electronic States of Single-Component Molecular Conductors [$M$(tmdt)$_2$]
}

\author{%
Hitoshi~\textsc{Seo}$^{1,2}$\thanks{E-mail address: seo@riken.jp},
Shoji~\textsc{Ishibashi}$^{3}$, 
Yuichi~\textsc{Otsuka}$^{4}$, 
Hidetoshi~\textsc{Fukuyama}$^{5,6}$, 
and~Kiyoyuki~\textsc{Terakura}$^{3,7}$
}

\inst{%
$^1$Condensed Matter Theory Laboratory, RIKEN, Wako, Saitama 351-0198 \\
$^2$JST, CREST, Saitama 351-0198 \\
$^3$Nanosystem Research Institute (NRI) ``RICS", AIST, Tsukuba, Ibaraki 305-8568 \\
$^4$Advanced Institute for Computational Science, RIKEN, Kobe, Hyogo 650-0047 \\
$^5$Department of Applied Physics, Tokyo University of Science, Shinjuku-ku, Tokyo 162-8601 \\
$^6$Research Institute for Science and Technology, Tokyo University of Science, Shinjuku-ku, Tokyo 162-8601 \\
$^7$JAIST, Nomi, Ishikawa 923-1292
}

\abst{
The electronic states of isostructural single-component molecular conductors 
 [$M$(tmdt)$_2$] ($M$~=~Ni, Au, and Cu) 
 are theoretically studied.
By considering fragments of molecular orbitals as basis functions, 
 we construct a multiorbital model common for the three materials. 
The tight-binding parameters are estimated 
 from results of first-principles band calculations, 
 leading to a systematic view of their electronic structures. 
We find that the interplay between 
 a $p\pi$-type orbital (L) on each of the two ligands  
 and a $pd\sigma$-type orbital (M$\sigma$) centered on the metal site 
 plays a crucial role: 
 their energy difference controls the electronic states near the Fermi energy.  
For the magnetic materials ($M$~=~Au and Cu), 
 we take into account Coulomb interactions on different orbitals, 
 i.e., we consider the multiorbital Hubbard model. 
Its ground-state properties are calculated within mean-field approximation 
 where various types of magnetic structures with different orbital natures are found. 
An explanation for the experimental results in [Cu(tmdt)$_2$] is provided:
 The quasi-degeneracy of the two types of orbitals leads to a dual state   
 where localized M$\sigma$ spins appear, 
 and L sites show a nonmagnetic state owing to dimerization. 
On the other hand, 
  [Au(tmdt)$_2$] locates in the subtle region in terms of the degree of orbital mixing. 
We propose possible scenarios for its puzzling antiferromagnetic phase transition, 
 involving the M$\sigma$ orbital 
 in contrast to previous discussions mostly concentrating on the L sector. 
}

\kword{molecular conductors, single-component molecular metal, multiorbital system, 
Hubbard model, first-principles band calculation, magnetism, metal-insulator transition, 
Mott insulator}

\begin{document}
\maketitle

\section{Introduction} 
\label{sec1}

Molecular crystals composed of one molecular species showing electrical conduction, 
 i.e., single-component molecular conductors (SCMC)~\cite{Tanaka_2001Science,Kobayashi_2004CR}, 
 have been revealed to bear novel electronic states. 
In particular, their multiorbital nature has been recognized, 
 which is different from the situation in conventional charge transfer salts (CTS), 
 where, in most cases, only one molecular orbital (MO) 
 contributes to their electronic properties~\cite{Seo_2004CR,Review_2006JPSJ}. 
In fact, the involvement of different MO  
 is a consequence of the molecular design~\cite{Kobayashi_2001JMC} 
 for realizing SCMC: 
 to make the energy difference between the frontier MO small enough, 
 so that their energy bands can overlap
 when inter-molecular transfer integrals become sufficiently large. 
Metal complex molecules of the form $M$($L$)$_2$ ($M$ = metal, $L$ = ligand) 
 are suitable for this purpose. 
Their frontier MO are 
 approximately bonding and antibonding combinations of 
 the $p\pi$ wave functions from the two ligands. 
Large ligands lead to an effectively small transfer integral between them 
 and result in a small energy difference. 
Such a situation indicates a two-MO system, 
 which is also realized in some CTS 
 as notably discussed in $M$(dmit)$_2$-based 
 compounds.~\cite{Canadell_1989JP,RKato_2004CR} 

The first SCMC in which metallic conductivity was reported~\cite{Tanaka_2001Science}
 is [Ni(tmdt)$_2$] 
 (tmdt = trimethylene-tetrathiafulvalene-dithiolate),~\cite{noteSqBr}
 whose resistivity decreases by cooling down to lowest temperatures ($T$). 
Direct evidence of its metallic feature was given by the observation of 
 three-dimensional Fermi surfaces by de Haas-van Alphen oscillations~\cite{Tanaka_2004JACS}, 
 whose results are consistent with first-principles band calculations~\cite{Tanaka_2004JACS,Rovira_2002PRB}.  
Near the Fermi energy $\epsilon_{\rm F}$, 
 there exist two overlapping bands from different $p\pi$-type MO 
 and $\epsilon_{\rm F}$ crosses the overlapping area. 
Electron and hole pockets appear, 
 owing to the existence of an even number of electrons in the unit cell 
 consisting of one Ni(tmdt)$_2$ molecule. 
This is the success of the molecular design mentioned above. 
It shows Pauli paramagnetic susceptibility,~\cite{Tanaka_2001Science} 
 while isotropic magnetoresistance suggesting the spin effect is observed~\cite{Yasuzuka_2008JPSJ}
 whose origin remains unclear. 

Since the discovery of [Ni(tmdt)$_2$], 
 many related compounds have been synthesized. 
Among them, an isostructural analog 
 but with an odd number of electrons per unit cell, 
[Au(tmdt)$_2$]~\cite{Suzuki_2003JACS},  
 has been attracting interest. 
It shows an antiferromagnetic (AF) phase transition with 
 a transition temperature \tn\ $=110$~K~\cite{Zhou_2006JACS,Hara_2008JPSJ}, 
 which is exceptionally high among molecular conductors. 
An intriguing point is that in the resistivity, 
 showing a metallic $T$-dependence down to low $T$ as well,
 no anomaly at around \tn\ is found~\cite{Tanaka_2007CL}. 
Furthermore, 
 the analysis of an NMR measurement~\cite{Hara_2008JPSJ} suggests 
 the magnetic moment in this AF state to be rather large, i.e., on the order of 1~$\mu_{\rm B}$. 
These features are 
 distinct from the formation of a spin-density-wave state 
 due to the nesting of Fermi surface, 
 as frequently observed in CTS, 
 where anomalies in transport properties appear 
 and typical values of the magnetic moment 
 are one order of magnitude smaller, 
 or even less~\cite{Review_2006JPSJ}. 
Such a magnetic solution attributed to the $p\pi$ bands 
 is actually stabilized in first-principles calculations~\cite{Ishibashi_2005JPSJ,Ishibashi_2008JPSJ}
 as well as in a mean-field (MF) study of an effective Hubbard model~\cite{Seo_2008JPSJ}, 
 which faces difficulties in explaining these experimental facts. 

Recently, another isostructural member [Cu(tmdt)$_2$], 
 having an odd number of electrons per unit cell, 
 has been successfully synthesized~\cite{Zhou_2010IC}. 
It shows a semiconductive behavior in contrast to the two compounds above, 
 and exhibits an AF phase transition  
 at \tn\ $=13$~K, much lower than in [Au(tmdt)$_2$]~\cite{Zhou_2010IC,Takagi_2012PRB}. 
The $T$ dependence of magnetic susceptibility above \tn \ is ascribed to 
 the behavior of the one-dimensional (1D) spin-$1/2$ Heisenberg model with AF exchange coupling of about 150~meV~\cite{Zhou_2010IC}, 
 which is consistent with 
 the $^1$H-NMR nuclear spin-lattice relaxation rate ($T_1^{-1}$) 
 indicating 1D spin dynamics~\cite{Takagi_2012PRB}. 
In this compound, in contrast with the discussions above, 
 a $pd\sigma$-type MO centered at the metal site is suggested to 
 lie close to the two ligands $p\pi$ orbitals, 
 and mix substantially. 
The charge transfer from the $p\pi$-MO 
 results in a nearly half-filled $pd\sigma$-band~\cite{Ishibashi_2012Crystals}. 
Then, the magnetic properties of [Cu(tmdt)$_2$] are attributed  
 to localized spins appearing on the $pd\sigma$-MO~\cite{Zhou_2010IC,Takagi_2012PRB}. 

In fact, the possibility that more than the two $p\pi$-MO  
 are involved in the electronic states of SCMC 
 was first proposed for [Au(tmdt)$_2$] on the basis of
 first-principles band calculations~\cite{Ishibashi_2005JPSJ,Ishibashi_2008JPSJ}: 
The $pd\sigma$- and $p\pi$-MO mix slightly 
 when forming the electronic band structure, 
 whereas the latter plays the major role near $\epsilon_{\rm F}$.  
However, more recently, 
 it has been inferred from experiments 
 that the orbital energy difference between these MO 
 is modified upon cooling by an unusual structural variation, 
 enhancing the mixing~\cite{Zhou_2009IC}.

Such multi-MO characters in SCMC 
 can be captured by the effective model approach 
 based on tight-binding approximation, 
 which has been successful in describing the electronic properties of CTS and 
 is now widely used~\cite{Seo_2004CR,Review_2006JPSJ,Seo_2006JPSJ}.
The observed de Haas-van Alphen oscillations in [Ni(tmdt)$_2$] are consistent 
 with the tight-binding picture~\cite{Tanaka_2004JACS}.  
In ref.~\citen{Seo_2008JPSJ}, we proposed that the basis sets for 
 the effective models of 
 [Ni(tmdt)$_2$] and [Au(tmdt)$_2$] 
 can be taken as virtual orbitals whose wave functions are 
 parts of the relevant MO near  $\epsilon_{\rm F}$, 
 rather than the MO themselves. 
In this paper, we 
 extend our theoretical approach to the newly synthesized [Cu(tmdt)$_2$] and 
 seek for a systematic view of the electronic states 
 among the isostructural family of [$M$(tmdt)$_2$] ($M$ = Ni, Au, and Cu).

In \S~\ref{sec2}, we set up our effective 
 multiorbital Hubbard model 
 and derive tight-binding parameters 
 by fitting the results of first-principles band calculations. 
By considering a common set of basis functions for the three materials, 
 a systematic view of the electronic states is achieved. 
Essentially, 
 the transfer integrals providing the structures of each band are similar among the members, 
 and orbital mixing is mostly governed by the energy difference between 
 the $p\pi$- and $pd\sigma$-type orbitals. 

Then, in \S~\ref{sec3}, by treating the on-site Coulomb interactions 
 within MF approximation, we investigate the ground-state properties 
 of models for [Au(tmdt)$_2$] and [Cu(tmdt)$_2$]. 
We will see that orbital mixing brings about 
 phase diagrams showing different magnetic states
 when Coulomb interactions on the two types of orbitals are independently varied. 
In particular, a slight enhancement of mixing in [Au(tmdt)$_2$] 
 suggested by experiments~\cite{Zhou_2009IC} 
 results in marked changes from our previous results~\cite{Seo_2008JPSJ}: 
The involvement of the $pd\sigma$ orbital is suggested. 

Section~\ref{sec4} is devoted to discussions, 
 especially on the magnetic transitions observed in the two compounds. 
Our results are consistent with the picture that, 
 in [Cu(tmdt)$_2$], 
 the $pd\sigma$-MO carries 1D $S$~= 1/2 localized spins, 
 interpreted as a multiband Mott insulator. 
We discuss possible situations realized in [Au(tmdt)$_2$],  
 from the viewpoint of doped Mott insulating systems due to orbital mixing. 
A summary is given in \S~\ref{sec5}. 


\section{Effective Model}\label{sec2}
The wave functions 
 that we chose as basis sets for the effective model of SCMC in our previous work~\cite{Seo_2008JPSJ} 
 are localized on some portions of the molecules. 
They can be considered as fragments of the MO, 
 and then called the fragment MO (fMO) 
 in refs.~\citen{Bonnet_2010JCP,Tsuchiizu_2011JPSJ,Tsuchiizu_2012JCP}, 
 which we follow in this paper as well. 

The original motivation to consider such decomposition of MO 
 was the results of first-principles calculations~\cite{Ishibashi_2005JPSJ,Ishibashi_2008JPSJ}. 
The spin-dependent calculation for [Au(tmdt)$_2$] indicates 
 a stable AF pattern where spins align oppositely within each molecule. 
To understand this unusual situation, 
 it was speculated that a ligand $p\pi$-$p\pi$ transfer integral 
 is larger between adjacent molecules 
 than that within a molecule, 
 which is supported by the results of an analysis based on the fMO approach~\cite{Seo_2008JPSJ}.
These features are consistent with the molecular design~\cite{Kobayashi_2001JMC} 
 mentioned in \S~\ref{sec1} and imply that the fMO approach gives an insightful picture
 of the electronic states of SCMC.
Such discussions have recently been elaborated 
 within quantum chemistry calculations~\cite{Bonnet_2010JCP,Tsuchiizu_2011JPSJ,Tsuchiizu_2012JCP}, 
 for (TTM-TTP)I$_3$ and [Au(tmdt)$_2$]. 
It is shown that the two approaches, the use of the MO and fMO pictures, 
 can be transformed from one to the other.  
We note that the fMO approach here and the so-called fragment molecular
 orbital method applied to huge molecules~\cite{Kitaura_1999CPL,Tsuneyuki_2009CPL} 
 share common concepts.

In the following, 
 all first-principles calculations, 
 including those for MO of isolated molecules,  
 are performed using the computational code QMAS (Quantum MAterials Simulator)~\cite{QMAS} 
 based on the projector augmented-wave method~\cite{PAW} 
 with the generalized gradient approximation~\cite{GGA}. 
See refs.~\citen{Ishibashi_2008JPSJ} and \citen{Ishibashi_2012Crystals} 
 for details. 
\begin{figure}[b]
\centerline{\includegraphics[width=8.4truecm]{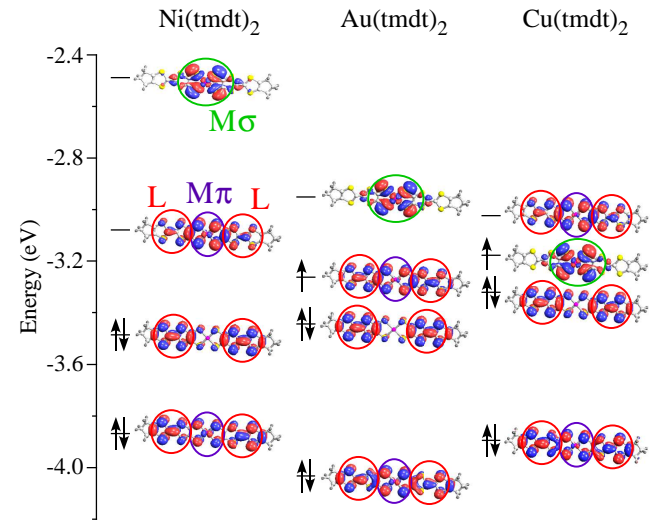}}
\vspace*{0em}
\caption{(Color online)
Molecular orbitals for $M$(tmdt)$_2$ ($M$ = Ni, Au, and Cu) molecules, 
 from which the energy bands in their crystals near the Fermi energy are 
 formed~\cite{Ishibashi_2005JPSJ,Ishibashi_2008JPSJ,Ishibashi_2012Crystals}. 
Nonmagnetic wave functions and energy eigenvalues together with the electron occupation 
 for isolated molecules are shown. 
The rough spatial extensions of the M$\sigma$, L, and M$\pi$ orbitals 
 explained in the text are indicated. 
}
\label{fig1}
\end{figure}

\subsection{Molecular orbitals and fragment model}\label{subsec2-1}
The band structures of isostructural SCMC near $\epsilon_{\rm F}$ are composed of 
 several MO with similar characters 
 upon chemical modifications~\cite{Kobayashi_2004CR,Ishibashi_2005JPSJ,Ishibashi_2008JPSJ}; 
 this applies to the family of [$M$(tmdt)$_2$] 
 including the new member $M$~=~Cu~\cite{Zhou_2010IC,Ishibashi_2012Crystals}. 
Four MO which mostly contribute to the 
 electronic bands near $\epsilon_{\rm F}$
 are shown in Fig.~\ref{fig1}.

They can be approximately reconstructed using three kinds of fMO, 
 which we call here M$\sigma$, L, and M$\pi$. 
The M$\sigma$ and M$\pi$ orbitals are the $p$-$d$ mixed wave functions, 
 roughly being 
 an anti-bonding combination of the metal site $d_{xy}$ and $d_{xz}$ orbitals, 
 and the surrounding S $2p$ orbitals, respectively~\cite{noteMnotation}. 
The relevant atomic $d$ orbitals are 3$d$ for $M$ = Ni and Cu, and 5$d$ for $M$ = Au. 
The L orbital is the $p\pi$ orbital 
 which is similar to the HOMO of the TTF molecule  
 embedded in the ligands (see ref.~\citen{Kobayashi_2004CR}). 
There are two of them in one molecule, i.e., 
 L1 and L2, one for each ligand; 
 they are equivalent due to the inversion center at the metal site. 

In isolated molecules, 
 the M$\sigma$ orbital does not mix with other orbitals from their symmetry;
 thus, it is a MO itself, 
 i.e., the $pd\sigma$-MO mentioned in \S~\ref{sec1}. 
As can be seen in Fig.~\ref{fig1}, 
 the other three MO 
 can roughly be described as linear combinations of L and M$\pi$ orbitals~\cite{Seo_2008JPSJ} as 
 $\phi_{\rm L1}-\phi_{\rm L2}+c_1\phi_{\textrm{M}\pi}$, $\phi_{\rm L1}+\phi_{\rm L2}$, and 
 $\phi_{\rm L1}-\phi_{\rm L2}-c_2\phi_{\textrm{M}\pi}$, 
 where $c_1$ and $c_2$ are some coefficients~\cite{Bonnet_2010JCP}  
(we omit renormalization factors). 

In the fMO scheme, 
 we consider these three kinds of orbitals 
 as a basis set composing the band structures, 
 and then for the low-energy effective model. 
The two-MO case mentioned in \S~1 
 corresponds to the situation where only the L orbitals are considered. 
In ref.~\citen{Seo_2008JPSJ}, we chose  
 \{L, M$\pi$\} for [Ni(tmdt)$_2$] 
 and \{M$\sigma$, L\} for [Au(tmdt)$_2$]  
 to reproduce the first-principles band structures near $\epsilon_{\rm F}$. 
Here, 
 all \{M$\sigma$, L, M$\pi$\} are taken into account as a common set, 
 in order to provide a systematic view of the compounds.

Our model Hamiltonian including local Coulomb interactions reads:%
\begin{align}
&{\cal H} = {\cal H}_0 + {\cal H}_{\rm int},\label{eq:H}\\
&{\cal H}_0 =\sum_{\langle l,m \rangle} \sum_s t_{lm} \left( c^\dagger_{ls} c_{ms}^{} + \mathrm{h.c.} \right) \nonumber\\
 &\hspace{3em} 
+ \sum_{i} \left( \Delta_{\textrm{M}\sigma} \ n_i^{\textrm{M}\sigma} 
 + \Delta_{\textrm{M}\pi} \ n_i^{\textrm{M}\pi} \right),\label{eq:H0}\\
&{\cal H}_{\rm int} = \sum_{i} \left\{ U_{\textrm{L}} 
    \left( n_{i\uparrow}^\textrm{L1} n_{i\downarrow}^\textrm{L1} +n_{i\uparrow}^\textrm{L2} n_{i\downarrow}^\textrm{L2} \right) \right. 
+ U_{\textrm{M}\sigma} \ n_{i\uparrow}^{\textrm{M}\sigma} n_{i\downarrow}^{\textrm{M}\sigma} \nonumber\\
 &\hspace{3em} 
+  U_{\textrm{M}\pi} \ n_{i\uparrow}^{\textrm{M}\pi} n_{i\downarrow}^{\textrm{M}\pi} 
 + \left. U'_\textrm{M} \ n_i^{\textrm{M}\sigma} n_i^{\textrm{M}\pi} \right\},\label{eq:Hint}
\end{align}
where ${\cal H}_0$ and ${\cal H}_{\rm int}$ represent 
 the one-particle part, determining the band structure, 
 and the on-site interaction, respectively.
In the first term of eq.~(\ref{eq:H0}), 
 $t_{lm}$ denotes the transfer integrals between fMO, 
 where the sum is taken for inter-fMO pairs $\langle l,m \rangle$ including 
 intra- and inter-molecular ones, 
 and $c_{ls}$ ($c^\dagger_{ls}$) 
 denotes the annihilation (creation) operator 
 for all kinds of orbitals with fMO site index $l$ and spin $s=\uparrow$ or $\downarrow$. 
In the second term of eq.~(\ref{eq:H0}), 
 $\Delta_{\textrm{M}\sigma}$ and $\Delta_{\textrm{M}\pi}$ are the orbital energies 
 of the M$\sigma$ and M$\pi$ orbitals, with respect to the L level. 
The sum here, as well as in eq.~(\ref{eq:Hint}), is taken for the molecule index $i$, 
 where the number operators are 
 $n_{is}^\textrm{o}={c^{\textrm{o}}_{is}}^\dagger c^{\textrm{o}}_{is}$ 
 and $n_{i}^\textrm{o}= n_{i\uparrow}^\textrm{o}+n_{i\downarrow}^\textrm{o}$ 
 with an orbital index $\textrm{o}=$ M$\sigma$, L1, L2, or M$\pi$. 
The intraorbital on-site Coulomb interactions for the three kinds of fMO are denoted as 
 $U_{\textrm{L}}$, $U_{\textrm{M}\sigma}$, and $U_{\textrm{M}\pi}$.
As for the interorbital on-site interaction, 
 we only include 
 $U'_\textrm{M}$ between M$\sigma$ and M$\pi$ for simplicity, 
 considering that these two orbitals 
 share the spatial extent 
 while they are separated from the L orbitals~\cite{noteHund}.

\begin{figure*}[htb]
\vspace*{1em}
\begin{center}
 \includegraphics[width=7.5cm,clip]{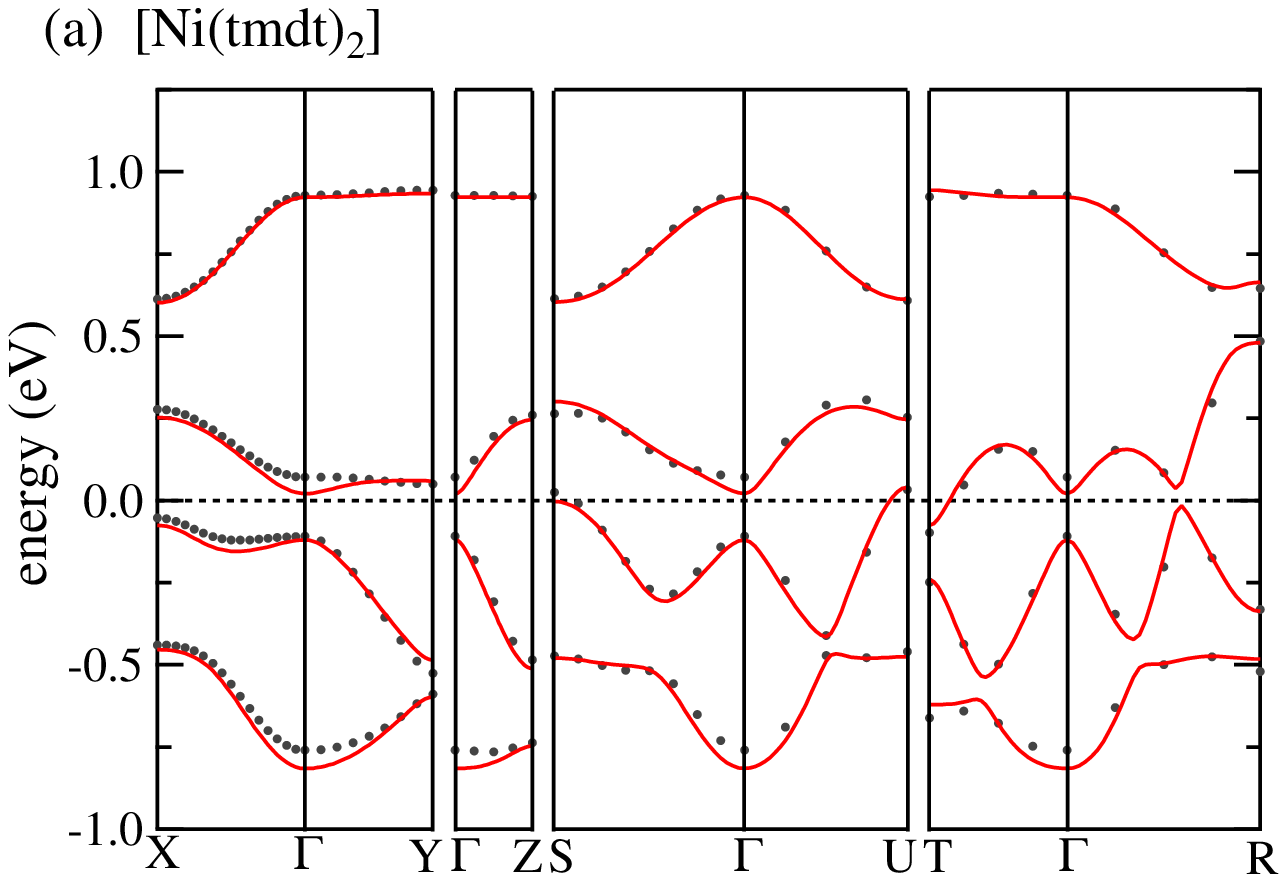}
 \includegraphics[width=7.5cm,clip]{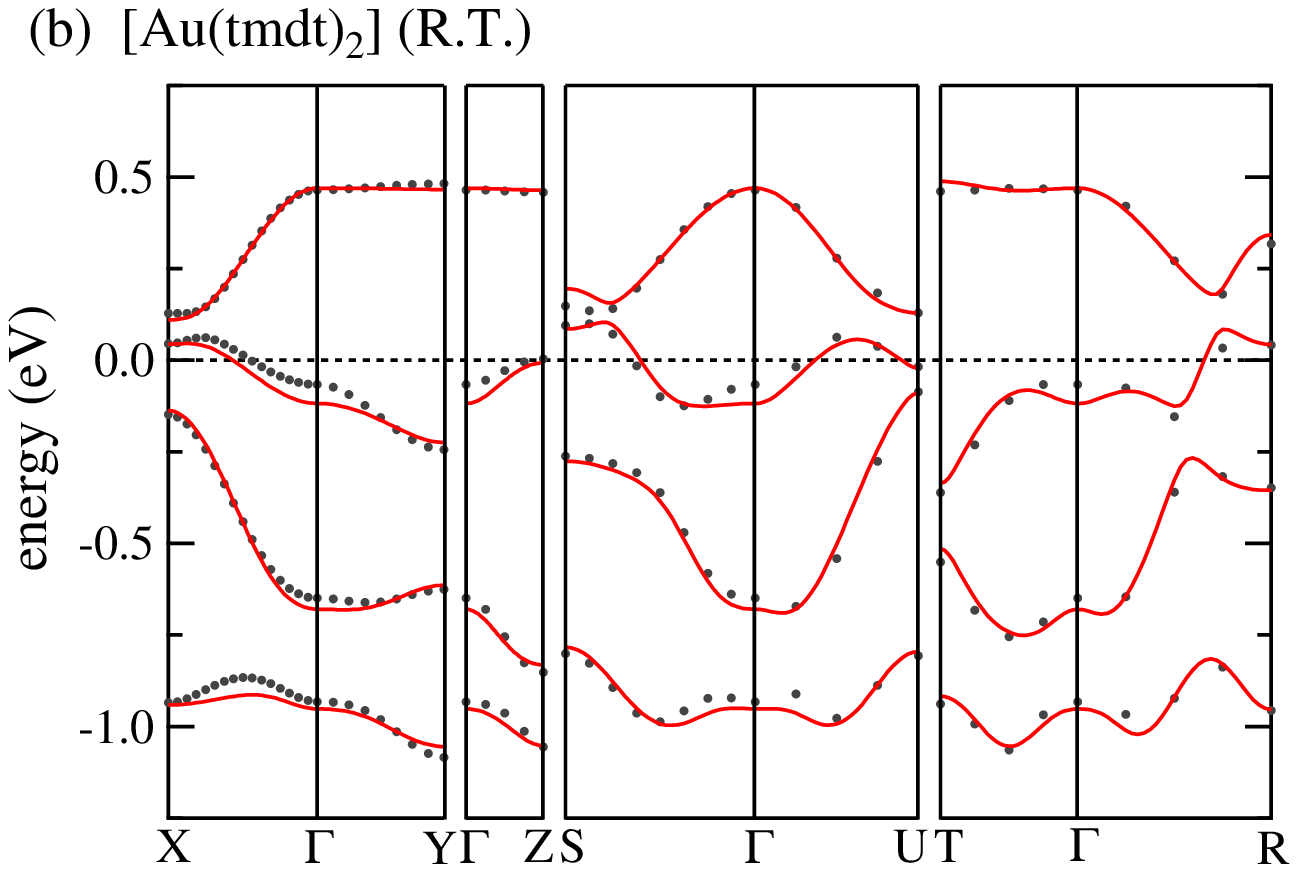}\\
 \includegraphics[width=7.5cm,clip]{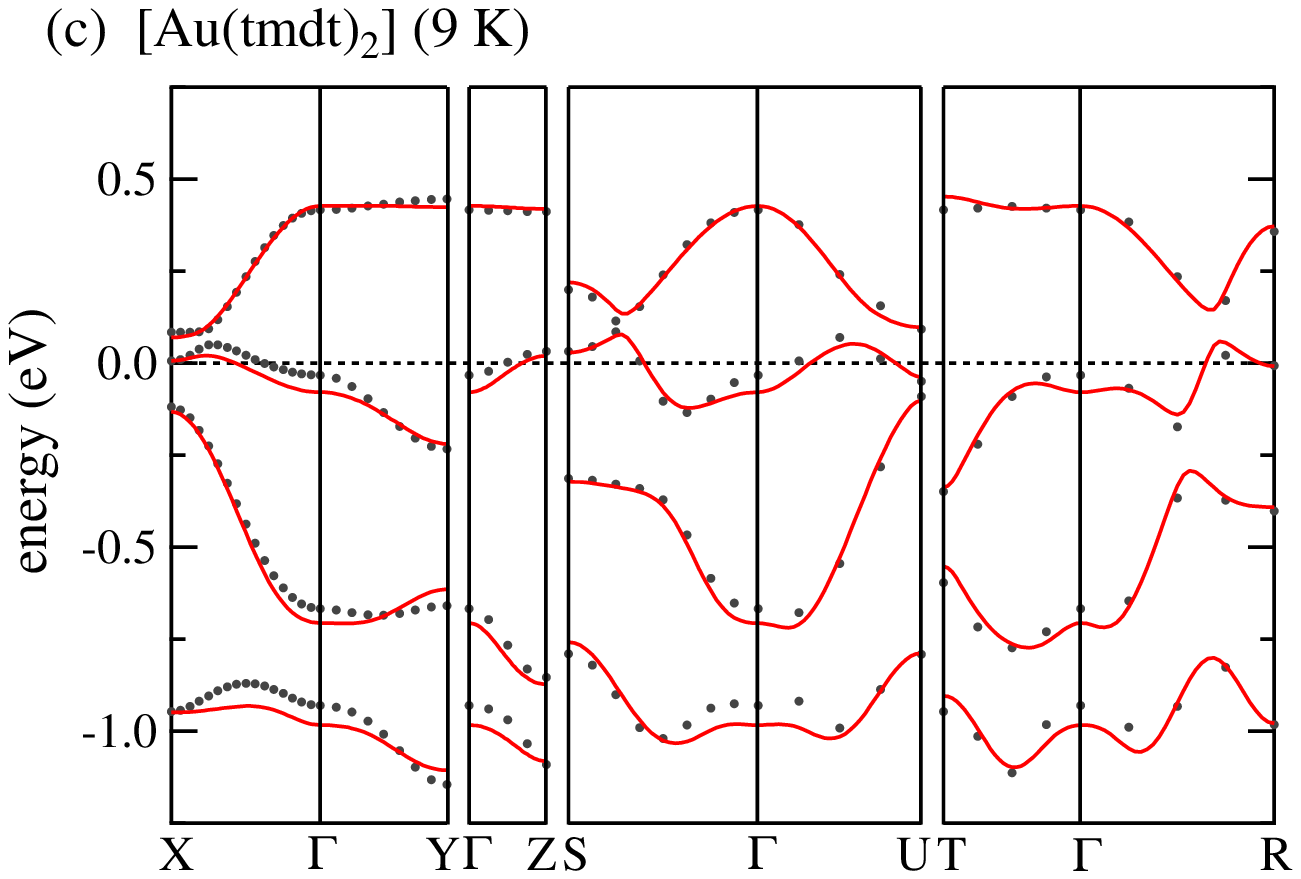}
 \includegraphics[width=7.5cm,clip]{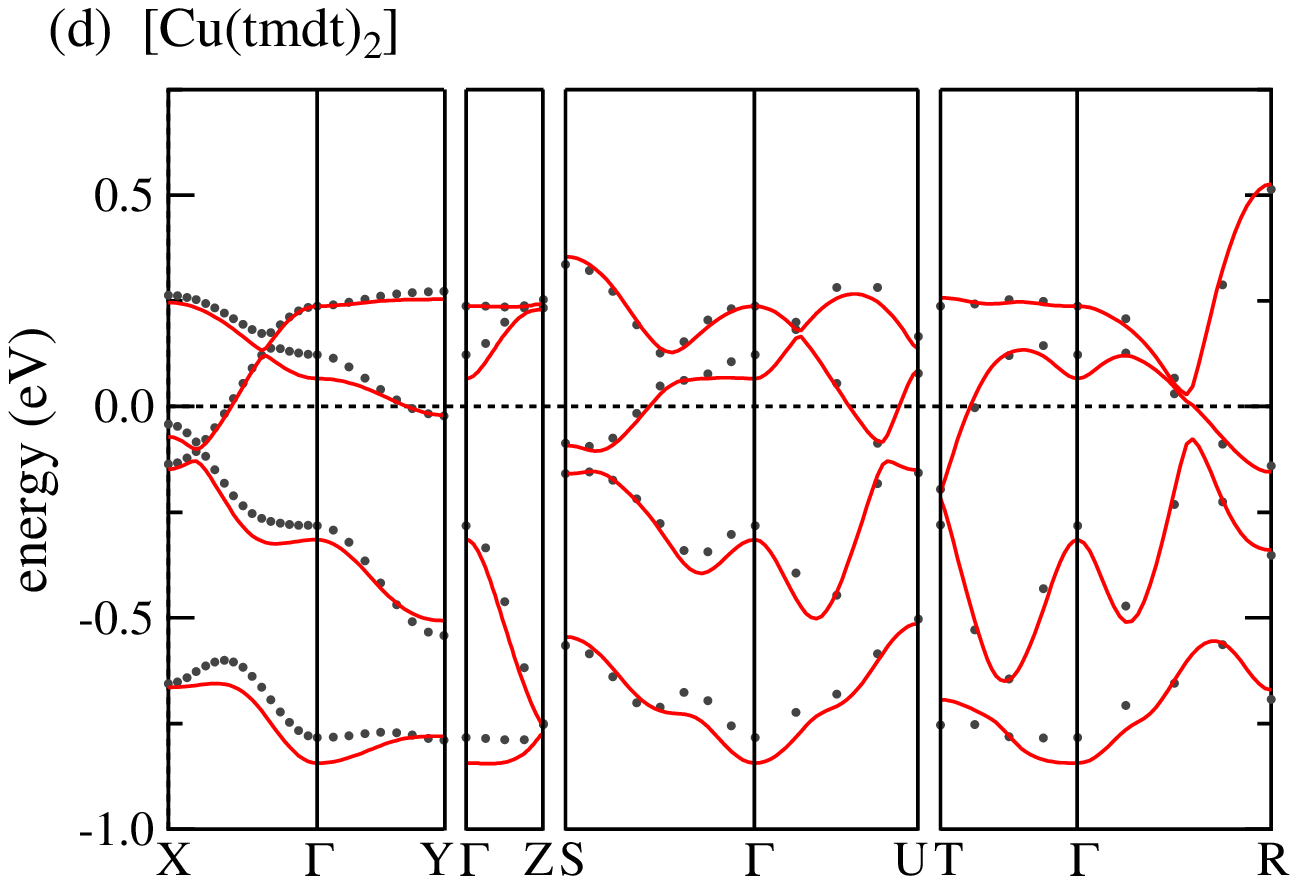}
\end{center} 
\vspace*{-1em}
\caption{(Color online)
Tight-binding dispersions (lines) 
 obtained by a fitting to first-principles band structure (dots)
 near the Fermi energy $\epsilon_{\rm F}$ (dotted lines) 
 of (a) [Ni(tmdt)$_2$], [Au(tmdt)$_2$]  for (b) room-temperature and  
(c) 9~K structures, 
 and (d) [Cu(tmdt)$_2$]. 
The dispersions are drawn 
 between the $\Gamma$-point, and X($\pi$,0,0), Y(0,$\pi$,0), Z(0,0,$\pi$), 
 S($\pi$,$\pi$,0), U($\pi$,0,$\pi$), T(0,$\pi$,$\pi$), and R($\pi$,$\pi$,$\pi$)-points, 
 and first-principles calculation data used for the fitting along them are plotted. 
}
\vspace*{-1.5em}
\label{fig2}
\end{figure*}%
\begin{table*}[htb]
\caption{Tight-binding parameters, $t_{lm}$, 
and orbital energies from L level, 
 $\Delta_{\textrm{M}\sigma}^0$, and $\Delta_{\textrm{M}\pi}^0$,
 obtained by a fitting to first-principles band calculations. 
For example, 
 bond: A, site pair: L1-M$\sigma$, 
 denotes $t_{lm}$ along the A ([1,0,0]) bond 
 between the L1 and M$\sigma$ sites 
 belonging to the molecule at the origin (0, 0, 0) and that at ($a$, 0, 0), 
 respectively. As for notations for intermolecular bonds, see Fig.~\ref{fig3}(a).
$t_{lm}$ with large absolute values are shown by bold characters. 
Orbital occupancies per site 
 $\bar{n}_{\textrm{o}}$ ($\textrm{o}=$ M$\sigma$, L, M$\pi$) 
 are also listed. 
}
 \begin{center}
 \begin{tabular}{cccccc}
bond & site pair & [Ni(tmdt)$_2$] & [Au(tmdt)$_2$] & [Au(tmdt)$_2$] (9~K) & [Cu(tmdt)$_2$] \\ \hline
intra-mol. & L1-L2 & -61 meV & -24 meV &  -27 meV  & -35 meV \\ 
A & L1-L1, L2-L2 & {\bf -86} & {\bf -99}  & {\bf -98} & {\bf -90}\\
B & L1-L2 & {\bf 221} & {\bf 207}  & {\bf 222} & {\bf 250} \\
Q & L1-L2 & {\bf 131} & {\bf 109} & {\bf 129} & {\bf 134}  \\
P & L1-L2 & 33 & 41 & 46 & 37 \\ \hline
A & L1-M$\sigma$, M$\sigma$-L2 & 37 & 33 & 41 & 14 \\
B & L1-M$\sigma$, M$\sigma$-L2 & 27 & 27 & 33 & 29 \\
C & L1-M$\sigma$, M$\sigma$-L2 & -50 & -25 & -29 & -24 \\
R & L1-M$\sigma$, M$\sigma$-L2 & 21 & 22 & 27 & 14 \\ \hline
intra-mol. & L1-M$\pi$, M$\pi$-L2 & {\bf -200}  & {\bf -138}  & {\bf -127} & {\bf -188} \\ 
B & L1-M$\pi$, M$\pi$-L2 & 0 & 30 & 35 & 20 \\
C & L1-M$\pi$, M$\pi$-L2 & -22 & -5 & -22 & -7 \\
Q & L1-M$\pi$, M$\pi$-L2 & 5 & 15 & 14 & 11 \\ \hline
A & M$\sigma$-M$\sigma$ & {\bf 82} & {\bf 95} & {\bf 98} & {\bf 96} \\ 
A & M$\pi$-M$\pi$ & -13 & -23 & -30 & -19 \\ 
B & M$\pi$-M$\pi$ & 24 & 37 & 44 & 50 \\ 
A & M$\pi$-M$\sigma$ & 53 & -34 & -14 & -1 \\ \hline
\ \\
orbital energies \\ \hline
$\Delta_{\textrm{M}\sigma}^0$ & & 879 meV & 522 meV &  476 meV &  181 meV \\
$\Delta_{\textrm{M}\pi}^0$ & & -193  & -612  & -636  &  -377 \\ \hline
\ \\
orbital occupancies \\ \hline
$\bar{n}_{\textrm{M}\sigma}$ & & 0.02 & 0.03 & 0.09 & 0.86 \\
$\bar{n}_{\textrm{L}}$ & & 1.16  & 1.50 &  1.47 &  1.15 \\
$\bar{n}_{\textrm{M}\pi}$ & & 1.66 & 1.97 & 1.97 & 1.85 \\
\hline
 \end{tabular}
 \end{center}
\vspace*{-1.5em}
\label{table1}
\end{table*}

\subsection{Fitting to first-principles band calculations}\label{subsec2-2}
The tight-binding parameters are obtained by a numerical fitting 
 to first-principles band structures for the nonmagnetic state. 
In Fig.~\ref{fig2}, we show the bands near $\epsilon_{\rm F}$, 
 together with the fitted tight-binding dispersions. 
The four bands originate from the four MO, or equivalently, the four fMO; 
 The unit cell consists of one molecule. 
As for [Au(tmdt)$_2$], calculations were performed 
 for both room-$T$ and low-$T$ (9~K) structures determined experimentally, 
 due to the indication of a structural variation upon cooling~\cite{Zhou_2009IC}, 
 as noted above. 
In the calculations for [Ni(tmdt)$_2$] and [Cu(tmdt)$_2$], 
 the room-$T$ structure parameters are used. 

One can see that
 the top band in [Ni(tmdt)$_2$] 
 and the bottom band in [Au(tmdt)$_2$]
 are separated from the others. 
This is the reason we previously used three-band fits (two kinds of fMO)~\cite{Seo_2008JPSJ}. 
On the other hand, in [Cu(tmdt)$_2$], all four bands are overlapping, 
 requiring a four-band fit for a reasonable agreement with the first-principles band structure.
Note that the total band widths of the four bands are about 
 1.7~eV ($M=$~Ni) $>$ 1.6~eV (Au) $>$ 1.3~eV (Cu). 

The fitted tight-binding parameters 
 together with the orbital occupancies per site 
 calculated from ${\cal H}_0$, i.e., 
 $\langle n_i^{\textrm{o}} \rangle \equiv
 \bar{n}_{\textrm{o}}$ ($\textrm{o}=$ M$\sigma$, L, M$\pi$), 
 are listed in Table~\ref{table1}. 
The listed orbital energies, 
 $\Delta_{\textrm{M}\sigma}^0$, and $\Delta_{\textrm{M}\pi}^0$, 
 are obtained by fitting the energy dispersions
 of ${\cal H}_0$. 
By noting that the first-principles band structures are obtained 
 self-consistently including the Hartree contributions within the interactions, 
 we can make a correspondence between the fitted values 
 and the orbital energies in eq.~(\ref{eq:H}) as~\cite{Seo_2008JPSJ,Tsuchiizu_2012JCP,Misawa_2011JPSJ} 
\begin{align}
\Delta_{\textrm{M}\sigma}^0 &=
 \Delta_{\textrm{M}\sigma} 
 +U_{\textrm{M}\sigma} \bar{n}_{\textrm{M}\sigma}/2 
 +U'_{\textrm{M}} \bar{n}_{\textrm{M}\pi} 
 -U_{\textrm{L}} \bar{n}_{\textrm{L}}/2,\label{eq:orbene1}\\ 
\Delta_{\textrm{M}\pi}^0 &=
 \Delta_{\textrm{M}\pi} 
 +U_{\textrm{M}\pi} \bar{n}_{\textrm{M}\pi}/2 
 +U'_{\textrm{M}} \bar{n}_{\textrm{M}\sigma} 
 -U_{\textrm{L}} \bar{n}_{\textrm{L}}/2.\label{eq:orbene2}
\end{align}

\begin{figure}[htb]
 \centerline{\includegraphics[width=6cm]{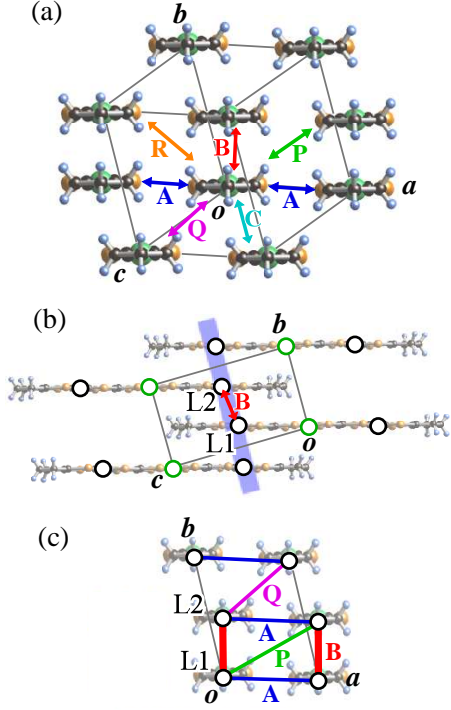}}
\caption{(Color online)
[$M$(tmdt)$_2$] viewed along the molecular long axis (a) 
 and schematic views of fMO model with lattice sites represented as circles [(b) and (c)]. 
The unit cell containing one molecule with four orbitals is shown as gray lines, 
 and notations for intermolecular bonds are indicated 
 whose correspondences are 
 A [100], B [111], C [101], P [211], Q [001], and R [011]. 
The two-dimensional network of L sites is shown 
 in (b), perpendicular to the plane, and in (c); 
 the sites connected by the largest transfer integral along the B bonds form dimers.  
See also Fig.~1 of ref.~\citen{Seo_2008JPSJ}.}
\label{fig3}
\end{figure}

The transfer integrals $t_{lm}$ 
 show more or less similar values for all three members~\cite{noteTransvalue}, 
 which is due to the fact that they are isostructural, 
 as noted in our previous work~\cite{Seo_2008JPSJ}. 
The M$\sigma$ orbitals have a large $t_{lm}$ only for the A bonds along the [100] direction: 
 they show a 1D structure. 
The L orbitals, on the other hand, possess a two-dimensional network, 
 where the {\it inter}-molecular dimers are formed by B bonds. 
Their network is schematically shown in Figs.~\ref{fig3}(b) and (c);  
 in the unit cell, the two L sites from different molecules 
 form L1-L2 dimers. 
The degree of dimerization, 
 represented by the intradimer transfer integral,
 namely, that along the L1-L2 B bond,  
 is largest in [Cu(tmdt)$_2$]. 
This dimerized structure resembles the situation commonly seen in typical CTS, 
 e.g., in TM$_2X$ and polytypes of ET$_2X$ such as the $\kappa$ and $\beta$-types~\cite{Review_2006JPSJ,Seo_2004CR}. 
M$\pi$ orbitals, in contrast, 
 do not have large $t_{lm}$ between them, 
 but bridge L layers mainly along the intramolecular bonds. 
They are appreciable (0.1~-~0.2~eV) 
 and then these two orbitals mix with each other. 
The main differences between the three compounds 
 are in orbital energy, 
 which we will discuss in the next subsection. 

\subsection{Systematic view of electronic structures}\label{subsec2-3}

\begin{figure*}[htb]
\vspace*{1em}
 \centerline{\includegraphics[width=14cm,clip]{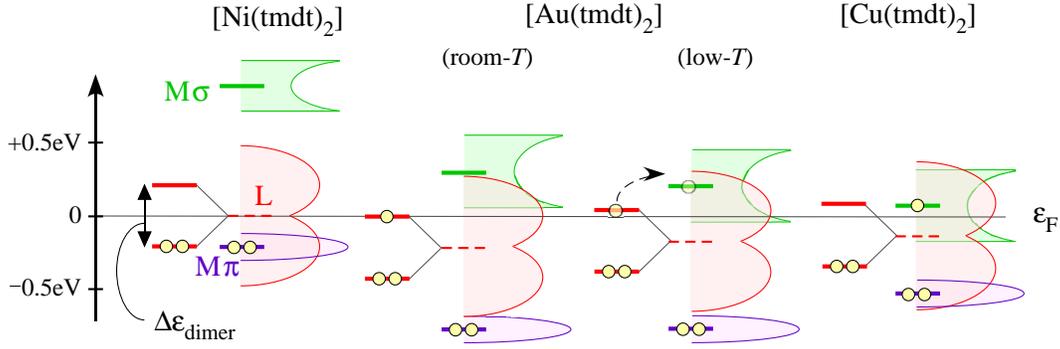}}
\vspace*{0em}
\caption{(Color)
Energy diagram of orbital energies based on Table~\ref{table1}, 
 with respect to Fermi energy $\epsilon_\textrm{F}$, 
 together with approximate electron occupancies: 
 4 electrons for $M$~= Ni and 5 electrons for $M$~= Au and Cu in total, 
 in the unit cell of one molecule having four orbitals. 
The degenerated L level (broken red line) splits due to intermolecular dimerization;  
 $\Delta\epsilon_\textrm{dimer}$ represents the dimerization gap. 
The schematic density of states (DOS) is also shown for each orbital: 
 M$\sigma$ has a one-dimensional electronic structure, and 
 the dimerization in L is respresented by a two-hump structure. 
Note that the orbital mixing is not considered here. 
(See refs.~\citen{Ishibashi_2005JPSJ,Ishibashi_2008JPSJ,Ishibashi_2012Crystals} 
 for the actual DOS by first-principles calculation.)
}
\vspace*{-2em}
\label{fig4}
\end{figure*}

The orbital energies $\Delta_{\textrm{M}\sigma}^0$ and $\Delta_{\textrm{M}\pi}^0$ 
 together with the transfer integrals determining the band structures, 
 and the electron numbers, 
 4 ($M$~=~Ni) or 5 ($M$~=~Au, Cu) in the four orbitals,  
 lead us to a systematic view of this family, 
 as shown in Fig.~\ref{fig4}.

The M$\pi$ level is low in energy 
 but mixes with the L bands particularly in [Ni(tmdt)$_2$] and [Cu(tmdt)$_2$]; 
 $\bar{n}_{\textrm{M}\pi}$ deviates from 2, 
 as seen in Table~\ref{table1}. 
However, 
 as far as the main characteristics of the electronic states near $\epsilon_{\rm F}$ are concerned, 
 the M$\pi$ orbital does not play an important role. 
Then, 
 $\Delta_{\textrm{M}\sigma}^0$ is the essential difference
 among the three members, 
 controlling the orbital mixing between the M$\sigma$ and L orbitals. 
It becomes monotonically small as 
 $M$ = Ni $\rightarrow$ Au~(room-$T$) $\rightarrow$ Au~(low-$T$) $\rightarrow$ Cu. 
This gives rise to a crucial difference in the magnetic states of the Au and Cu systems,
 as will be shown in \S~3.

The M$\sigma$ level in [Ni(tmdt)$_2$] is high, 
 and approximately 2 electrons enters the L level. 
The existence of two L sites in the unit cell, 
 which show dimerization as discussed in the previous subsection, 
 results in band splitting, 
 but not large enough to generate a direct band gap. 
Then electron and hole pockets appear and compose the Fermi surface.
 This is how the first SCMC with a metallic character was realized. 

Comparing the room-$T$ and low-$T$ structure parameters for [Au(tmdt)$_2$], 
 the low-$T$ data show slightly larger transfer integrals 
 than the room-$T$ data especially in L-L pairs; 
 this is naturally expected from the thermal contraction. 
$\Delta_{\textrm{M}\sigma}^0$ is reduced by lowering $T$ by about 0.05~eV, 
 which is consistent with the previous MO calculation~\cite{Zhou_2009IC}. 
As a result, 
 a fraction of electrons in the L sites, 
 nearly 3/4-filled in the room-$T$ parameters, 
 are transfered to the M$\sigma$ orbital in the low-$T$ parameters: 
\{$\bar{n}_{\textrm{M}\sigma}$, $\bar{n}_{\textrm{L}}$\} = \{0.03, 1.5\} (room-$T$) 
$\rightarrow$ \{0.09, 1.47\} (low-$T$). 

As for [Cu(tmdt)$_2$], 
 M$\sigma$ mixes more with L 
 due to the further reduction in $\Delta_{\textrm{M}\sigma}^0$, 
 by about 0.30 eV smaller than the low-$T$ parameter for [Au(tmdt)$_2$].   
In particular, near $\epsilon_{\rm F}$, 
 the contribution of M$\sigma$ is appreciable~\cite{Ishibashi_2012Crystals}, 
 even though the orbital level scheme shows 
 a positive $\Delta_{\textrm{M}\sigma}^0$. 
This is because L orbitals have larger $t_{lm}$ with a two-dimensional character, 
 while the M$\sigma$ band is 1D; 
 therefore, the former show wider bands. 
The orbital occupancies are \{$\bar{n}_{\textrm{M}\sigma}$, $\bar{n}_{\textrm{L}}$\} = \{0.86, 1.15\}, 
 which are substantially varied from the cases above. 
These features are close to the situation in the MO scheme in Fig.~\ref{fig1}: 
In Cu(tmdt)$_2$, 
 the M$\sigma$ ($pd\sigma$) orbital is occupied with nearly one electron. 

Let us comment on the relation between our level scheme 
 and the nominal charges given as Ni$^{2+}$[(tmdt)$^-$]$_2$, 
 Au$^{3+}$[(tmdt)$^{1.5-}$]$_2$, and Cu$^{2+}$[(tmdt)$^-$]$_2$.
Their corresponding occupations of metal $d$ levels 
 are Ni:(3$d$)$^8$, Au:(5$d$)$^8$, and Cu:(3$d$)$^9$, respectively. 
When one hypothesizes that the M$\sigma$ and M$\pi$ levels are `$d$' levels of the metal atoms 
 and the L orbitals for the full charge of the ligands 
 (namely, omit $p$-$d$ mixing, which is actually large), 
 these nominal charges correspond to, 
 (M$\pi$)$^2$[(L)$^1$]$_2$,  (M$\pi$)$^2$[(L)$^{1.5}$]$_2$, 
 and (M$\pi$)$^2$(M$\sigma$)$^1$[(L)$^1$]$_2$, respectively, 
namely, a full-filled M$\pi$ in all compounds, 
and, 
(i) in $M$~=~Ni, a 1/2-filled L band, 
(ii) in $M$~=~Au, a 3/4-filled L band, 
and (iii) in $M$~=~Cu, a 1/2-filled M$\sigma$ band and a 1/2-filled L band. 
These are close to the situations in Fig.~\ref{fig4}. 

\section{Mean-Field Calculation}\label{sec3}

\begin{figure}
\vspace*{1em}
 \centerline{\includegraphics[width=7.2cm]{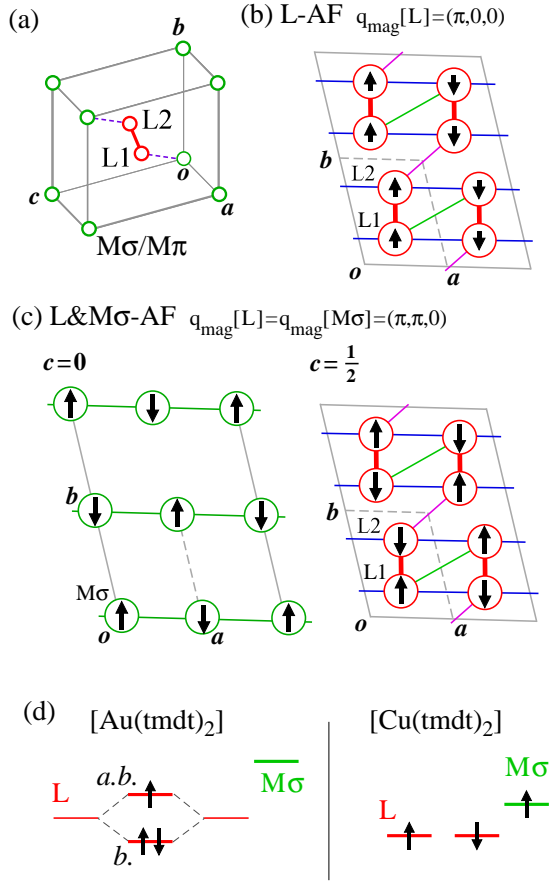}}
\vspace*{0em}
\caption{(Color)
Schematic views of (a) unit cell containing four orbitals 
 with the red bar indicating the inter-molecular B bond in Fig. 3, 
 and (b) and (c) representative antiferromagnetic (AF) solutions 
 projected on the $ab$ plane [see Fig.~\ref{fig3}(c)]. 
(b) L-AF: spin moments appear only on L sites [dimer-AF pattern with 
 spin ordering vector $q_\textrm{mag}[\textrm{L}]=(\pi,0,0)$] 
 seen for parameters of [Au(tmdt)$_2$]~\cite{Ishibashi_2005JPSJ,Seo_2008JPSJ}.
(c) L\&M$\sigma$-AF: spins align in staggered AF manner among L and M$\sigma$ sites 
 for [Cu(tmdt)$_2$] with common ordering vector 
 $q_\textrm{mag}[\textrm{L}]=q_\textrm{mag}[\textrm{M}\sigma]=(\pi,\pi,0)$. 
In (d), 
 schematic representations of electron occupancies
 realized in parameters for [Au(tmdt)$_2$] and [Cu(tmdt)$_2$] are shown, 
 respectively. 
 `{\it b}.' and `{\it a}. {\it b}.' represent the bonding and antibonding orbitals 
 for an L dimer, respectively. 
} 
\vspace*{-1.5em}
\label{fig5}
\end{figure}

As mentioned in \S~\ref{sec1}, 
 [Au(tmdt)$_2$] and [Cu(tmdt)$_2$] show phase transitions to magnetically ordered states. 
Here, we study their ground-state ($T$ = 0) magnetic states  
 where the interaction terms in ${\cal H}_{\rm int}$ [eq.~(\ref{eq:Hint})] are treated 
 by MF approximation as  
 $n_{is}^\textrm{o} n_{is'}^\textrm{o'} \rightarrow 
 \langle n_{is}^\textrm{o} \rangle n_{is'}^\textrm{o'} 
 + n_{is}^\textrm{o} \langle n_{is'}^\textrm{o'} \rangle
 - \langle n_{is}^\textrm{o} \rangle \langle n_{is'}^\textrm{o'}\rangle$. 
Such a MF treatment is suitable in seeking for possible different states, 
 as in our model here with multiple degrees of freedom. 
In the calculations, 
 we consider a supercell of $2a \times 2b \times 2c$, 
 which includes \{M$\sigma$, L (L1 and L2), M$\pi$\} $\times$~8 orbitals [see Fig.~\ref{fig5}(a)].%

We set the values of $t_{lm}$ to the fitted results in Table~\ref{table1}, 
 while $\Delta_{\textrm{M}\sigma}$ and $\Delta_{\textrm{M}\pi}$ are adjusted 
 by the conditions in eqs.~(\ref{eq:orbene1}) and (\ref{eq:orbene2}) 
 when the interaction parameters are varied. 
Then the band dispersions are unchanged within the paramagnetic metallic (PM) solution. 
Namely, 
 the orbital occupations in the PM state, 
 $\langle n_{i\uparrow}^\textrm{o} \rangle = \langle n_{i\downarrow}^\textrm{o} \rangle =
 \langle n_i^\textrm{o} \rangle/2$, 
 are fixed at the values listed in Table~\ref{table1} as 
 $ \langle n_i^\textrm{o} \rangle= \bar{n}_{\textrm{o}}$. 
As for the magnetic solutions, 
 we relax the condition of fixed occupation 
 and searched for self-consistent solutions of the lowest energy 
 in an unrestricted manner within the periodicity of the supercell. 

In the following, 
 we set $U_{\textrm{M}\sigma} = U_{\textrm{M}\pi} = U'_{\textrm{M}} \equiv U_{\textrm{M}}$ 
 for simplicity~\cite{noteUparam} and vary $U_{\textrm{M}}$ and $U_{\textrm{L}}$ independently.
In the whole parameter range we sought, 
 the M$\pi$ orbital has a negligible spin moment, i.e., it is magnetically inactive. 
Then the parameters $U_{\textrm{M}}$ and $U_{\textrm{L}}$ control 
 the correlation effect on the M$\sigma$ and L orbitals, respectively. 
One speculation we can make is the relation $U_{\textrm{M}} \gsim U_{\textrm{L}}$, 
 considering the $d$ contribution to the M$\sigma$ orbital 
 as well as its smaller spatial extent than the L orbital. 

Before presenting the results, 
 we remark about two typical AF solutions. 
The first is stabilized in the case of [Au(tmdt)$_2$] when $U_{\textrm{L}}$ is increased, 
 whose nature was  
 discussed in refs.~\citen{Ishibashi_2005JPSJ} and \citen{Seo_2008JPSJ}. 
As shown in Fig.~\ref{fig5}(b), 
 spin moments appear only on L sites; therefore, this state is represented as L-AF. 
The spins are parallel within L1-L2 dimers connected by B bonds [see Fig.~\ref{fig3}(c)] 
 and antiparallel between dimers along interdimer bonds
 denoted as A, P, and Q in Fig.~3(a).
 Their ordering vector is $q_\textrm{mag}[\textrm{L}]=(\pi,0,0)$. 
This pattern corresponds to the dimer-AF spin order 
 frequently appearing in 1/4-filled CTS under dimerization~\cite{Seo_2004CR}. 
When the on-site Coulomb repulsion is sufficiently large, 
 their staggered pattern can open a gap at $\epsilon_{\rm F}$ 
 when the system is 1/4-filled in terms of either electrons or holes, 
 with each dimer carrying an effective $S=$~1/2. 
Such a case is considered as the dimer-Mott insulating state. 
In the case of the room-$T$ parameters for [Au(tmdt)$_2$], 
 L bands are 3/4-filled, without pronounced mixing with other orbitals.  
Dimers are formed by intermolecular fMO sites, 
 and spins on two L sites in a molecule become antiparallel~\cite{Ishibashi_2005JPSJ}. 
The results of the first-principles calculation~\cite{Ishibashi_2005JPSJ} 
 correspond to the case of relatively small $U_\textrm{L}$ , 
 and the system remains metallic even in the L-AF state.

On the other hand, 
 in the case of [Cu(tmdt)$_2$],
 a stable AF pattern shows 
 spin moments on both L and M$\sigma$ sites
 when both $U_{\textrm{M}}$ and $U_{\textrm{L}}$ are large enough; 
 it is then denoted as L\&M$\sigma$-AF. 
As shown in Fig.~\ref{fig5}(c), 
 spins on L sites show a staggered AF state, 
 in which those connected with A, B, and Q (see Fig.~\ref{fig3}) bonds are antiparallel; 
 the $t_{lm}$ values are large along these L-L bonds, as shown in Table~\ref{table1}. 
Spins on M$\sigma$ sites are also staggered, 
 with the common spin ordering vector $q_\textrm{mag}[\textrm{L}]=q_\textrm{mag}[\textrm{M}\sigma]=(\pi,\pi,0)$. 
This can open an insulating gap when both L and M$\sigma$ bands becomes nearly 1/2-filled; 
 both orbitals provide effective $S=$~1/2, and 
 the AF pattern corresponds to the N{\' e}el state configuration. 
Now, this can be considered as a multiband Mott insulator, 
 due to the quasi-degeneracy of the two orbitals. 

The two contrasting situations for large interactions 
 are summarized in Fig.~\ref{fig5}(d), 
 where schematic representations of electron occupancies in the two cases are shown. 
In the case of [Au(tmdt)$_2$], localized spins appear on the antibonding orbital of dimerized L sites. 
In contrast, in the case of [Cu(tmdt)$_2$] 
 spins appear on each of the L sites as well as of the M$\sigma$ sites 
 as a result of orbital mixing. 
\begin{figure}
\begin{center}
\includegraphics[width=5.8cm,clip]{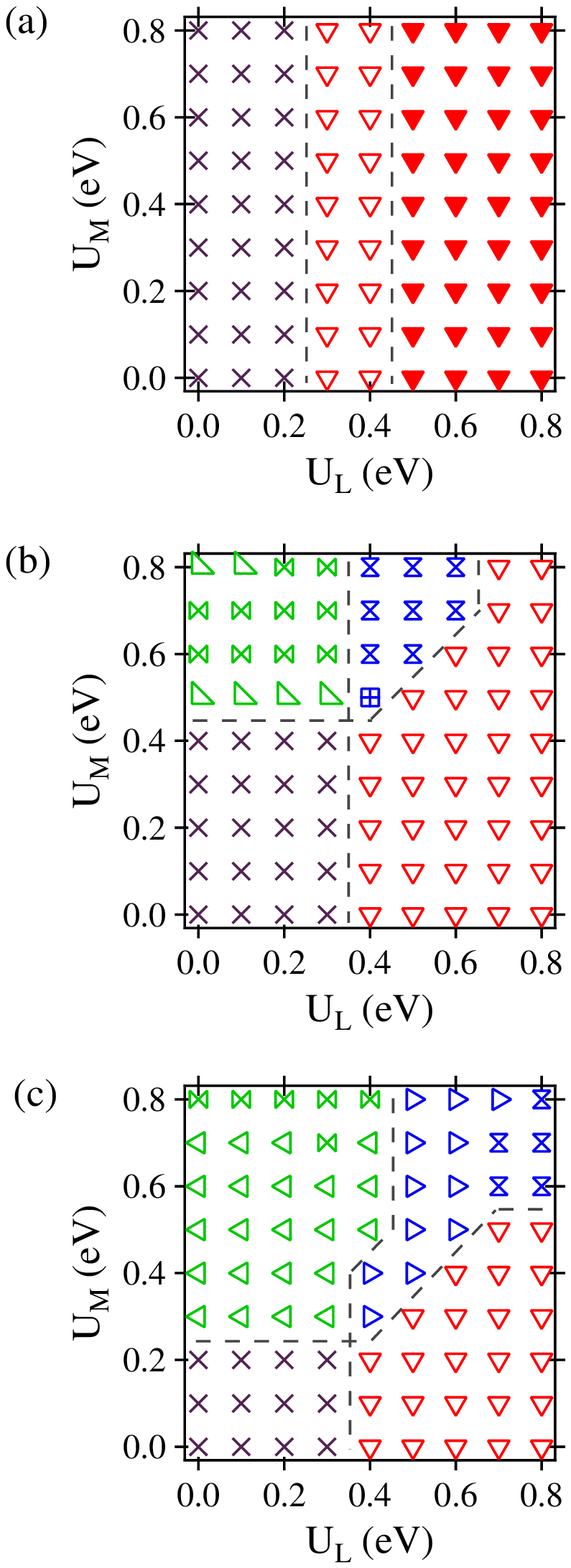}\\
\vspace*{0.5em}
\includegraphics[width=5cm,clip]{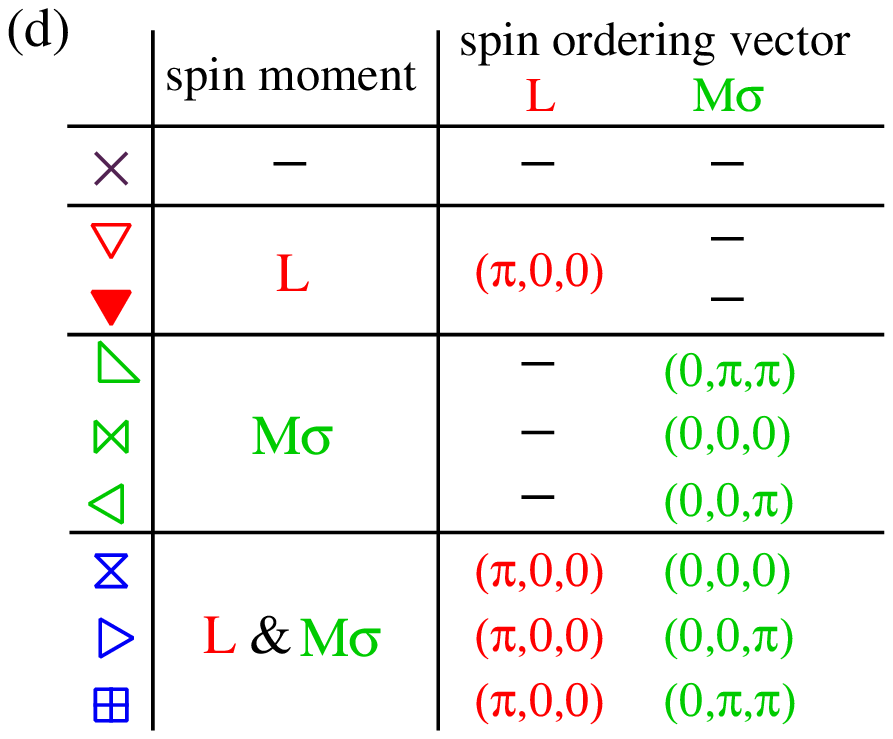}
\end{center}
\vspace*{0em}
\caption{(Color)
Mean field ground-state phase diagrams for [Au(tmdt)$_2$].
The parameters used are  
fitted results for (a) room-temperature ($T$) structure, 
(b) low-$T$ structure, 
and (c) room-$T$ structure but with reduced 
$\Delta_{\textrm{M}\sigma}^0 \rightarrow \Delta_{\textrm{M}\sigma}^0 - 0.1$~eV. 
Dashed lines are guides for the eyes.
The legend symbols are shown in (d); colors are appointed for 
distinctions between orbitals showing spin moments as 
red (L), green (M$\sigma$), and blue (both L and M$\sigma$), 
while filled symbols represent insulating states. 
}
\label{fig6}
\end{figure}
\begin{figure}
\vspace*{1em}
\begin{center}
\includegraphics[width=8.4cm,clip]{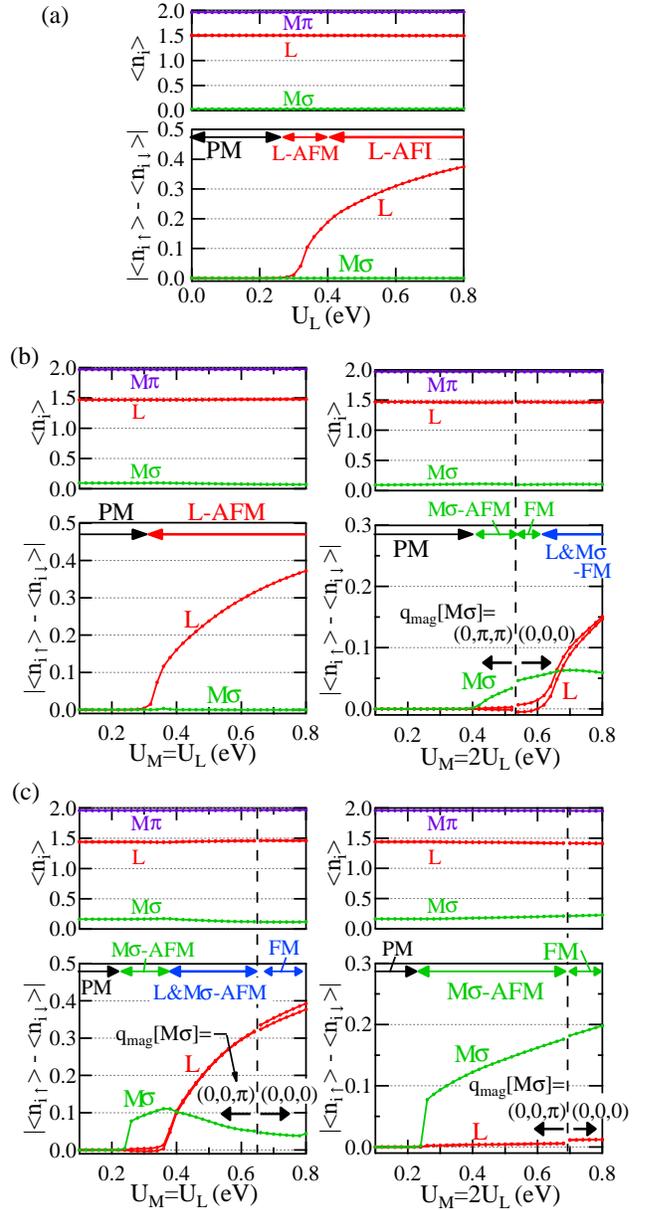}
\end{center}
\vspace*{-2em}
\caption{(Color)
Parameter dependences of expectation values of 
 electron density, $\langle n_i \rangle$, 
 and spin density, $|\langle n_{i\uparrow} \rangle - \langle n_{i\downarrow} \rangle|$,
 for each orbital site in lowest-energy MF solutions for Au(tmdt)$_2$. 
(a)-(c) correspond to the cases in Figs.~\ref{fig6}(a)-6(c), respectively. 
The left (right) panels in (b) and (c) are for a fixed ratio of 
 $U_{\textrm{M}}=U_{\textrm{L}}$ ($U_{\textrm{M}}=2U_{\textrm{L}}$). 
$q_\textrm{mag}[\textrm{M}\sigma]$ denotes the magnetic ordering vector for M$\sigma$ sites 
 that changes at the parameters indicated by the broken lines. 
($q_\textrm{mag}[\textrm{L}]$ is always $(\pi,0,0)$, as shown in Fig.~\ref{fig6}(d).)
}
\vspace*{-2em}
\label{fig7}
\end{figure}
\subsection{[Au(tmdt)$_2$]}\label{subsec31}

In Fig.~\ref{fig6}, 
 we show ground-state phase diagrams 
 for three different parameter sets corresponding to [Au(tmdt)$_2$]. 
Figures~\ref{fig6}(a) and \ref{fig6}(b) are those with 
 fitted results for the room-$T$ and low-$T$ structures,
 respectively, 
 in Table~\ref{table1}. 
Besides them, to see the effect of the reduction in $\Delta_{\textrm{M}\sigma}^0$ 
 more explicitly, 
 we artificially decrease it from its room-$T$ value 
 as $\Delta_{\textrm{M}\sigma}^0 \rightarrow \Delta_{\textrm{M}\sigma}^0-0.1$~eV, 
 while leaving the other room-$T$ parameters unchanged: 
 This is shown in Fig.~\ref{fig6}(c). 
The distinct states are indicated in the phase diagrams by different symbols
 summarized in Fig.~\ref{fig6}(d), 
 together with their magnetic ordering vectors.
As can be seen there, 
 the AF order on the L sites always has the pattern 
 shown in Fig.~\ref{fig5}(b) with $q_\textrm{mag}[\textrm{L}]=(\pi,0,0)$. 
By the reduction in $\Delta_{\textrm{M}\sigma}^0$, 
 we indeed find a crucial difference: 
The dimer-type AF insulating (AFI) state within the L orbital 
 discussed above [filled symbol in Fig.~\ref{fig6}(a)] 
 stabilized in the wide range of parameters in the room-$T$ case is not seen, 
 and different AF metallic (AFM) states appear in Figs.~\ref{fig6}(b) and \ref{fig6}(c), 
 owing to the mixing between the L and M$\sigma$ orbitals. 
 
In Fig.~\ref{fig7}, 
 we show the $U_{\textrm{L}}$ dependence 
 of site occupation number for the three kinds of orbitals
 together with 
 the spin moments on the L and M$\sigma$ orbitals. 
Figures~\ref{fig7}(a)-7(c) correspond to the cases in 
 Figs.~\ref{fig6}(a)-6(c), respectively. 
The right and left panels in Figs.~\ref{fig7}(b) and 7(c)
 are data along different traces in the ($U_{\textrm{M}}$, $U_{\textrm{L}}$) plane, 
 i.e., 
 for $U_{\textrm{M}}=U_{\textrm{L}}$ and $U_{\textrm{M}}=2U_{\textrm{L}}$, 
 respectively.%

\subsubsection{Room-temperature structure}\label{subsubsec311}
The phase diagram in Fig.~\ref{fig6}(a) shows no dependence 
 on $U_{\textrm{M}}$; 
 this is because the M$\sigma$ orbital is always nearly unoccupied, 
 as shown in Fig.~\ref{fig7}(a). 
As $U_{\textrm{L}}$ is increased, 
 the system varies as PM $\rightarrow$ AFM $\rightarrow$ AFI states. 
Since only L sites possess magnetic moments, the magnetic phases are represented as 
L-AFM and L-AFI in Fig.~\ref{fig7}(a). 

These results are almost identical to the results in ref.~\citen{Seo_2008JPSJ}, 
 where the MF calculations were carried out 
 for the three-band model based on \{M$\sigma$, L\}  
 and assuming $U_{\textrm{M}\sigma}=U_{\textrm{L}}$. 
This is consistent with the fact that the present results show 
 an almost fully occupied M$\pi$ orbital playing no role 
 and having no dependence on $U_{\textrm{M}}$. 
As discussed there, 
 the properties are governed by L orbitals. 
Their occupation number of 1.5 
 together with the rather strong dimerization 
 are consistent with the fact that 
 our results are analogous to the MF calculations 
 on two-dimensional 1/4-filled Hubbard models with dimerization, 
 e.g., as firstly performed on the model of $\kappa$-ET$_2X$\cite{Kino_1995JPSJ}. 
The existence of the AFM phase in between the AFI and PM phases 
 is due to the imperfect nesting property of the Fermi surface, 
 where small magnetic moments cannot produce a band splitting 
 large enough to open up a gap on the whole Fermi surface. 
As long as the system is in the AFM phase, 
 the spin moment on each site is less than 0.2~$\mu_\textrm{B}$; 
 this is also the same as that in the three-band model.~\cite{Seo_2008JPSJ}. 

\subsubsection{Low-temperature structure and reduced $\Delta_{\textrm{M}\sigma}^0$}\label{subsubsec312}
The phase diagrams in Fig.~\ref{fig6}(b) 
 for the fitted results for the low-$T$ structure  
 and in Fig.~\ref{fig6}(c) 
 for the room-$T$ values but with reduced $\Delta_{\textrm{M}\sigma}^0$ 
 share common features in their overall structure, 
 but are very different from that in the room-$T$ case. 
Spin moments appear on M$\sigma$ orbitals when $U_\textrm{M}$ is enlarged, 
 however insulating states are not stabilized, 
 at least, for $(U_\textrm{L},U_\textrm{M}) \leq 0.8$~eV; 
 the whole phase diagrams show metallic states. 
The phase diagrams are devided into four regions: 
 PM, L-AFM, 
 magnetic metallic states with moments on M$\sigma$ sites 
 [either antiferromagnetic (M$\sigma$-AFM) or ferromagnetic (M$\sigma$-FM)], 
 and those with moments on both orbitals 
 [L\&M$\sigma$-AFM or L\&M$\sigma$-FM]. 
These are common for Figs.~\ref{fig6}(b) and 6(c); 
 therefore, the main variation from the room-$T$ structure 
 to the low-$T$ structure can be captured 
 by the reduction in $\Delta_{\textrm{M}\sigma}^0$. 
We note that the reduction in $\Delta_{\textrm{M}\sigma}^0$ 
 when the room-$T$ and low-$T$ values are compared 
 is about 0.05 eV, while other parameters only slightly change; 
 the phase diagram shows a marked change.

The L-AFM phase stabilized 
 in the large-$U_\textrm{L}$, small-$U_\textrm{M}$ region 
 is the remnant of the L-AFI phase in the room-$T$ parameters. 
Its ordering vector is the same 
 and the amplitude of magnetic moment 
 is similar to that in the AFI phase in Fig.~\ref{fig6}(a). 
For example, in the left panel of Figs.~\ref{fig7}(b), 
 it reaches 0.37$\mu_\textrm{B}$ per site at $U_\textrm{M}=U_\textrm{L}=0.8$~eV.  
Nevertheless, the system does not turn into an insulating phase, 
 which is due to orbital mixing:
The occupation numbers 
 are $\langle n_{i}^{\textrm{M}\sigma} \rangle \simeq 0.1$
  and $\langle n_{i}^\textrm{L} \rangle \simeq 1.45$
for the low-$T$ parameters, 
 and $\langle n_{i}^{\textrm{M}\sigma} \rangle \simeq 0.2$ and  
  $\langle n_{i}^\textrm{L} \rangle \simeq 1.4$
 for the reduced-$\Delta_{\textrm{M}\sigma}^0$ case, 
 which are noticeably shifted from the case of the room-$T$ parameters. 
Then this L-AFM state in the large-$U_\textrm{L}$ region 
 can be considered as a `doped dimer-Mott insulator' 
 due to the mixing with the M$\sigma$ orbital. 

On the other hand, 
 in the small-$U_\textrm{L}$, large-$U_\textrm{M}$ region, 
 the M$\sigma$-AFM or M$\sigma$-FM state is stabilized. 
L sites remain paramagnetic and the system is naturally metallic. 
The magnitude of the moment appearing on M$\sigma$ sites is 
 limited by its occupation number, as shown in Figs.~\ref{fig7}(b) and 7(c), 
 i.e., about 0.1-0.2$\mu_\textrm{B}$.
It is difficult to discuss the origin of each spin ordering vector for M$\sigma$ sites, 
 $q_\textrm{mag}[\textrm{M}\sigma]$, 
 within our limited supercell size, 
 since its small filling factor would typically favor a longer periodicity. 
In fact, it takes different values delicately depending on the parameters. 
However, we consider that the region where spin moments appear is reasonable. 
Comparing the phase diagrams in Figs.~\ref{fig6}(b) and (c), 
 the latter has larger region of  phases with moments on the M$\sigma$ orbital, 
 owing to the smaller $\Delta_{\textrm{M}\sigma}^0$ 
 and therefore the larger occupation number in M$\sigma$. 

Even when we enter the region when both $U_\textrm{M}$ and $U_\textrm{L}$ are large 
 where spin ordering on M$\sigma$ and L sites coexist, 
 the system still remains metallic. 
Here, the spin moment on L sites can be large similarly to that in the case of the L-AFM phase. 
For example, in the left panel of Figs.~\ref{fig7}(c), 
 it reaches 0.35-0.4$\mu_\textrm{B}$ per site at 
 $U_\textrm{M}=U_\textrm{L}=0.8$~eV.  
In this sense, 
 this is also the remnant of the L-AFI phase
 in the room-$T$ parameters, 
 stabilized at similar interaction parameters. 
As shown in Fig.~\ref{fig7}(c), 
 when the spin moment on L sites increases, 
 that on the M$\sigma$ orbital decreases, 
 owing to the mismatch of their ordering vectors. 
This is in contrast with the case of Cu(tmdt)$_2$, 
 as we will see in the next subsection.

We note that, in the case of the low-$T$ parameters, 
 there is a wide region where the ordering vector is 
 $q_\textrm{mag}[\textrm{M}\sigma]=(0,0,0)$, 
 namely, the ferromagnetic (M$\sigma$-FM) 
 or ferrimagnetic state when L sites also show spin moments (L\&M$\sigma$-FM) 
 is stabilized. 
This can be ascribed to the large density of state at $\epsilon_{\rm F}$, 
 due to the lower edge of the 1D band 
 from the M$\sigma$ orbital~\cite{Ishibashi_2005JPSJ}. 
In the room-$T$ structure, this edge situates just above $\epsilon_{\rm F}$, 
 which becomes near it for the low-$T$ parameters. 
The states with $q_\textrm{mag}[\textrm{M}\sigma]=(0,0,0)$ also appears in 
 the reduced-$\Delta_{\textrm{M}\sigma}^0$ case
 [see Figs.~\ref{fig6}(c) and \ref{fig7}(c)]. 
  
\subsection{[Cu(tmdt)$_2$]}\label{subsec32}

\begin{figure}
\vspace*{2em}
\begin{center}
 \includegraphics[width=6cm,clip]{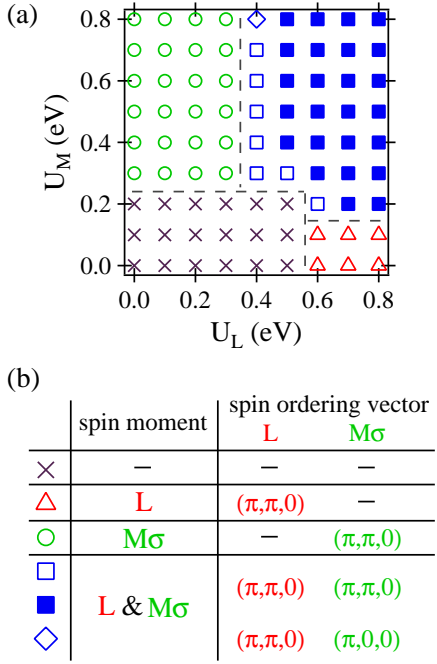}
\end{center}
\vspace*{0em}
\caption{(Color)
(a) Mean field ground-state phase diagram for [Cu(tmdt)$_2$]. 
The parameters used are 
the fitted results in Table~\ref{table1}. 
Dotted lines are guides for the eyes.
The legend symbols are shown in (b); 
colors are appointed for 
distinguishing between orbitals showing spin moments as 
red (L), green (M$\sigma$), and blue (both L and M$\sigma$), 
while filled symbols represent insulating states. 
}
\vspace*{-1.5em}
\label{fig8}
\end{figure}

\begin{figure}
\vspace*{2em}
\begin{center}
 \includegraphics[width=8.6cm,clip]{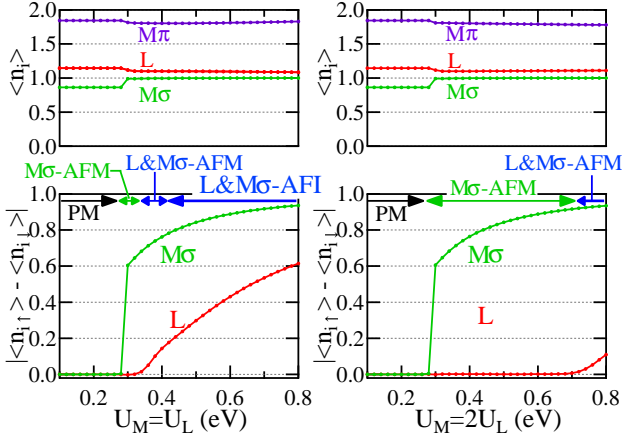}
\end{center}
\vspace*{-2em}
\caption{(Color)
Parameter dependences of expectation values of 
 electron density, $\langle n_i \rangle$, 
 and spin density, $|\langle n_{i\uparrow} \rangle - \langle n_{i\downarrow} \rangle|$,
 for each orbital site in the MF solutions for [Cu(tmdt)$_2$]. 
The left (right) panel is for fixed ratio of 
 $U_{\textrm{M}}=U_{\textrm{L}}$ ($U_{\textrm{M}}=2U_{\textrm{L}}$). 
}
\vspace*{-1.5em}
\label{fig9}
\end{figure}

The MF ground-state phase diagram for [Cu(tmdt)$_2$] is shown in Fig.~\ref{fig8},  
 while the parameter dependences of orbital occupations and magnetic moments  
 are shown in Fig.~\ref{fig9}. 
Although the M$\pi$ orbital is off from the full occupation compared with those in the other cases (see Table~\ref{table1}), 
 we find no contribution of it to the magnetic properties;
 therefore, we can set them aside again. 
The occupation numbers of the other two orbitals are now rather close to 1, i.e., 1/2-filling. 
Then, as discussed above, 
 an AF insulating state with both M$\sigma$ and L showing magnetic ordering 
 (L\&M$\sigma$-AFI) is seen to be stabilized 
 in the region where both $U_\textrm{L}$ and $U_\textrm{M}$ are large. 
Its spin pattern is the staggered one as is shown in Fig.~\ref{fig5}(c). 
As shown in Fig.~\ref{fig9}(c), 
 the spin moments on the L and M$\sigma$ orbitals 
 develop cooperatively as the interaction is enhanced.
%

In the large-$U_\textrm{L}$, small-$U_\textrm{M}$ region, 
 on the other hand, the L-AFM state is stabilized, 
 and vise versa, i.e., the M$\sigma$-AFM state is realized 
 for the small-$U_\textrm{L}$, large-$U_\textrm{M}$ region
In these states, 
 the spin ordering vector is the same as that in the L\&M$\sigma$-AFI phase 
 [$q_\textrm{mag}=(\pi,\pi,0)$]; 
 these states are continuously connected. 
However, they cannot open an insulating gap since, 
 although the AF ordering provides a band splitting at $\epsilon_{\rm F}$ 
 in the magnetic orbital sector, 
 the remaining orbital is paramagnetic with a Fermi surface. 
In other words, the spin ordering and band gap opening are orbital-selective. 
The critical value for the magnetic states 
 is smaller for the M$\sigma$-AFM, 
 consistent with the fact that M$\sigma$ sites  
 have smaller transfer integrals with a 1D structure 
 than the L sites. 
As shown in Fig.~\ref{fig9}, 
 the magnetic moment on each site can be large, 
 owing to the large filling factor near 1. 
In fact, the calculations show a jump in the charge density on each site 
 across the PM $\leftrightarrow$ AFM boundary
 (first order phase transition), 
 which results in the occupation number for M$\sigma$ and L 
 to be closer to 1 in the magnetic phases, 
 that is, 0.99 for M$\sigma$
 and 1.08 for L, 
 than in the PM phase. 

One point to note is that, 
 although the phase diagram is mostly dominated by the $q_\textrm{mag}=(\pi,\pi,0)$ state, 
 we find many self-consistent solutions with very close MF energies 
 in the region where M$\sigma$ orbitals are magnetic. 
In these quasi-degenerate states, 
 spin ordering vectors are of the form $q_\textrm{mag}[\textrm{M}\sigma]=(\pi,*,*)$
 with $*=0$ or $\pi$, namely, 
 only the 2$a$ periodicity is robust. 
This suggests that the spin exchange coupling between moments on M$\sigma$ sites 
 is essentially 1D, 
 which is consistent with the fitted results in Table~\ref{table1} 
 where the A bond along [100] has the largest M$\sigma$-M$\sigma$ transfer integral. 
This is also consistent with the first-principles calculations 
  that compare different magnetic orderings~\cite{Ishibashi_2012Crystals}. 
On the other hand, the L orbital always orders with $q_\textrm{mag}[\textrm{L}]=(\pi,\pi,0)$: 
 the two-dimensional AF is stable at the MF level.


\section{Discussion}\label{sec4}

In this section, 
 we compare our MF results with experiments on magnetic ordering in this family, 
 keeping in mind that, in the calculations, quantum fluctuations are neglected; 
 additional effects of such fluctuations as well as the possible strong correlation effect 
 can be speculated on top of the MF results. 
Discussions on [Cu(tmdt)$_2$] is followed by that on [Au(tmdt)$_2$], 
 since in the former we can have a consistent explanation of the experiments. 
As for the latter, 
 we present several possible scenarios for the magnetic phase transition in this compound, 
 in light of considerations of its Cu analog. 

\subsection{[Cu(tmdt)$_2$]}
Let us give an explanation of the experimental results in [Cu(tmdt)$_2$] 
 based on our calculations, 
 in association with the discussions 
 in the literature~\cite{Zhou_2010IC,Takagi_2012PRB,Ishibashi_2012Crystals}. 
The experiments show an insulating behavior in the resistivity, 
 and the magnetic properties of a 1D Heisenberg spin $S=1/2$ system 
 likely due to $pd\sigma$-MO, namely, the M$\sigma$ orbital.  

\begin{figure}[t]
\vspace*{2em}
 \centerline{\includegraphics[width=8.4cm,clip]{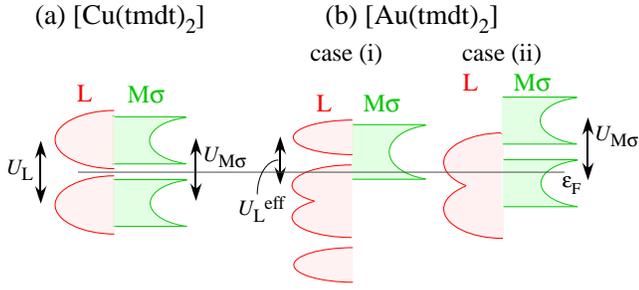}}
\vspace*{0em}
\caption{(Color online) 
Schematic density of states (DOS) for possible situations 
 for (a) [Cu(tmdt)$_2$] and (b) [Au(tmdt)$_2$]. 
The M$\pi$ orbital is omitted. 
The band splittings compared with the situations in Fig.~\ref{fig4} 
 are due to the formation of Mott gaps, 
 driven by $U_{\textrm{M}\sigma}$ for M$\sigma$ orbitals, 
 while for L orbitals driven by (a) $U_\textrm{L}$ and 
 (b) the effective on-dimer Coulomb repulsion indicated by $U_\textrm{L}^\textrm{eff}$. 
Coherent peaks expected in such doped Mott insulating systems are omitted as well. 
}
\vspace*{-1em}
\label{fig10}
\end{figure}

The insulating state in our MF calculations  
 is realized for the L\&M$\sigma$-AFI phase (Fig.~\ref{fig8})
 when both $U_\textrm{L}$ and $U_\textrm{M}$ are large. 
We can consider this state as a multiband Mott insulator  
 where both L and M$\sigma$ orbitals possess localized spins of effective $S=1/2$ each. 
In the MF calculation, this state is accompanied by a three-dimensional AF ordering. 
We can deduce the additional quantum effect as 
 a spin singlet formation in the L network: 
 the rather strong dimerization on B bonds can bring about a nonmagnetic ground state 
 in the L subunit, 
 as long as the other $t_{lm}$ values are small enough. 
Then the active spin degree of freedom arises only in the M$\sigma$ sector. 
It has a 1D character as discussed above, 
 and therefore consistent with the Mott insulating behavior 
 with 1D $S=1/2$ chains. 
This situation is schematically shown in Fig.~\ref{fig10}(a).  

Another scenario is that the system corresponds to 
 the M$\sigma$-AFM phase in our MF phase diagram 
 in the small-$U_\textrm{L}$ and large-$U_\textrm{M}$ region 
 (considering the relation $U_{\textrm{M}} \gsim U_{\textrm{L}}$), 
 while other effects beyond our calculation bring about the insulating behavior. 
In this MF solution, roughly speaking, 
 the magnetic ordering 
 brings about a gap at $\epsilon_\textrm{F}$ for the M$\sigma$ band, 
 but the L bands remains metallic. 
This corresponds to the state found in the first-principles band calculation~\cite{Ishibashi_2012Crystals}.
One possibility is that, 
 since, in this state, $\epsilon_\textrm{F}$ locates in the middle of L bands, 
 where the band overlap is small, a small perturbation might bring about a band gap. 
This is now a band insulator due to the dimerization in the L sector. 
Then the spin degree of freedom is only from the M$\sigma$ orbital, 
 again considered as a Mott insulator, 
 showing the same magnetic behavior as above. 
The largest degree of dimerization in L sites among the cases 
 listed in Table~\ref{table1} is consistent with both pictures. 

%

\subsection{[Au(tmdt)$_2$]}
There have been puzzling experimental data for [Au(tmdt)$_2$], 
 as mentioned in \S~\ref{sec1}. 
Our MF results for the room-$T$ structure in \S~\ref{subsubsec311}, 
 which suggest that only L orbitals are magnetically active, 
 lead to the same conclusion 
 that we discussed previously~\cite{Ishibashi_2005JPSJ,Ishibashi_2008JPSJ,Seo_2008JPSJ}, 
 facing difficulties in explaining the experimental results. 
The main problems were as follows: 
 (1) The AFM (spin-density-wave) state in the calculations shows a small magnetic moment, 
 whereas when $U_{\textrm{L}}$ is increased to achieve a large moment state 
 the system enters the AFI phase, 
 i.e., the dimer-Mott insulator. 
The experimentally observed metallic ground state 
 with a large magnetic moment could not be reconciled. 
(2) Discussions only involving the L sector are incompatible with 
 the absence of a sign of phase transition in the resistivity at $T_\textrm{AF}$. 
Spin-density-wave states due to the nesting of the Fermi surface 
 should result in a change in transport properties. 

In clear contrast, on the other hand, 
 the results of MF calculations in \S~\ref{subsubsec312}, 
 using the low-$T$ parameters 
 [Figs.~\ref{fig6}(b) and~\ref{fig7}(b)] 
 as well as 
 the reduced $\Delta_{\textrm{M}\sigma}^0$ values 
 [Figs.~\ref{fig6}(c) and~\ref{fig7}(c)], 
 show possible AFM states with larger magnetic moments. 
In these states, 
 owing to the orbital mixing between L and M$\sigma$, 
 AF ordering cannot open a gap at $\epsilon_\textrm{F}$; 
 therefore, the system remains metallic. 
Such an involvement of multiorbitals 
 also leads to possible explanations for 
 the absence of anomaly at $T_\textrm{AF}$ in the resistivity, 
 suggesting that the magnetic and transport properties are carried by different degrees of freedom. 
Below, let us propose two possibilities on the basis of our results, 
 considering the strong correlation effect in addition. 
We ascribe the regions where magnetic moments arise in our calculations 
 as doped Mott insulating states, as discussed in \S~\ref{subsubsec312}. 

(i) \{L: doped Mott insulator, M$\sigma$: PM\}.
This corresponds to the L-AFM state in our calculations. 
The L orbital forms a dimer-Mott insulating state 
 but doped with holes, 
 which are provided from the M$\sigma$ orbital remaining in a PM state. 
The Mott gap is due to the effective on-dimer Coulomb interaction, 
 as indicated by $U_\textrm{L}^\textrm{eff}$ in Fig.~\ref{fig10}(b) [case (i)]. 
 
(ii) \{L: PM, M$\sigma$: doped Mott insulator\}.
In the MF calculation, in the large-$U_\textrm{M}$, small $U_\textrm{L}$ region,  
 M$\sigma$-AFM/FM states are stabilized. 
If a similar picture of the Mott insulating nature in scenario (i) is applied, 
 this results in a situation shown in Fig.~\ref{fig10}(b) [case (ii)]. 
The Mott gap formation in the M$\sigma$ sector 
 is adopted from the discussions on [Cu(tmdt)$_2$]. 

In both cases, 
 the doped Mott insulating character will generate magnetism, 
 while there exist carriers unchanged upon magnetic ordering. 
In the MF phases corresponding to case (i), 
 large spin moments appear on L sites, 
 whereas its ordering is expected to affect the transport properties 
 since L bands are mainly responsible for the conduction. 
On the other hand, case (ii) is favorable in the sense that 
 L bands are paramagnetic and carries the charge transport, 
 and then M$\sigma$ sites are responsible for magnetism; 
 such a picture has been proposed on the basis of experimental considerations~\cite{Takagi_private}. 
The weaker dimerization in L sites than in the Cu analog obtained 
 in our estimated transfer integrals 
 may be a factor for stabilizing such a metallic L system. 
However, 
 in our calculation the magnetic moment on M$\sigma$ 
 is small, limited by its occupation, i.e., the small electron-doping level. 
This is not in agreement with the argument. 
With enhanced mixing between the two levels, 
 approaching the situation in [Cu(tmdt)$_2$] 
 is expected to bring about a possible reconciliation with the experimental results, 
 although within our estimation of parameters 
 such a prominent mixing is not achieved.
Future works including that on the strong correlation effect 
 are needed for further investigating this possibility. 

Finally, the observed peculiarly high $T_\textrm{AF}$ 
 is difficult to discuss from our calculation only for the ground state, 
 and the evaluation of transition temperature at the MF level is 
 usually not reliable for comparison with experiments, 
 especially when the strong correlation is involved. 
The low dimensionality in both L and M$\sigma$ orbitals 
 when considering intraorbital transfer integrals 
 is apparently incompatible with such a high $T_\textrm{AF}$, 
 considering the fact that the energy scale is similar to those of other molecular conductors 
 showing lower transition temperatures for magnetic ordering in general. 
One possibility is that doped carriers in mixed orbitals effectively 
 make the spin-spin interaction large, 
 especially even in the $c$-direction where the original parameters $t_{lm}$ 
 among the L and M$\sigma$ orbitals are small, 
 making the system three-dimensionally coupled. 
Then the low-dimensionality embedded in each orbital 
 can be released to increase the critical temperature. 

\section{Summary}\label{sec5}

We have constructed effective models of single-component molecular conductors 
 [$M$(tmdt)$_2$] ($M$ = Ni, Au, and Cu) 
 showing a multiorbital nature. 
Tight-binding parameters are obtained by a fitting to first-principles band structures. 
The fragment molecular orbital picture leads us to a systematic view of this family: 
 the interplay between a characteristic anisotropic electronic network 
 and the orbital energy difference can tune electronic states using a different choice of $M$, 
 particularly that between the $pd\sigma$-type and $p\pi$-type orbitals. 
By taking into account the Coulomb interaction, 
 we discussed, for [Au(tmdt)$_2$] and [Cu(tmdt)$_2$], 
 the mean-field phase diagrams and magnetic solutions of our effective model. 
In the former compound, 
 we suggest that the mixing between the two orbitals 
 can play a key role in resolving their puzzling experimental results. 
On the other hand, 
 in the latter, 
 the existence of a multiorbital Mott insulator is suggested, 
 which is consistent with the experimental results.  
Both of these cases are distinctive examples of molecular systems. 

\section*{Acknowledgments}
We thank K. Kanoda, A. Kobayashi, H. Kobayashi, R. Takagi, and M. Tsuchiizu
 for discussions and suggestions. 
This work was supported by Grant-in-Aid for Scientific Research
(Nos. 20110003, 20110004, and 24108511) 
from MEXT.


\begin{thebibliography}{99}
\bibitem{noteSqBr}
 In this paper, [$M$(tmdt)$_2$] and $M$(tmdt)$_2$ are used to express 
 the crystals and molecules, respectively.
\bibitem{noteHund}
 We have considered the effect of Hund coupling in the MF calculations, 
 but did not find any qualitative difference in the parameter range we investigated ($\le 0.2 U'_{\rm M}$), 
 and therefore omitted it in this paper.
\bibitem{noteUparam}
 We have investigated different parameter sets such as 
 $U_{\textrm{M}\sigma} = U_{\textrm{M}\pi}=1.25 U'_{\textrm{M}}$, 
 but we did not find any qualitative difference. 
The M$\pi$ orbital is always nearly occupied, 
 and the $U'_{\textrm{M}}$-term does not play a role 
 in stabilizing different states.
\end{thebibliography}
\end{document}